LABOUR MONITORING IN PREGNANT WOMEN USING

PHONOCARDIOGRAPHY, ELECTROCARDIOGRAPHY AND

ELECTROMYOGRAPHY TECHNIQUE.

Thesis submitted in partial fulfilment
of the requirements for the degree of

*(Master of Science in **Programme** by Research
Or Doctor of Philosophy in **Programme**)*

by

ANUSHKA TIWARI

2020702020

anushka.tiwari@research.iiit.ac.in

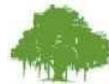

International Institute of Information Technology - Hyderabad

(Deemed to be University)

Hyderabad – 500032, INDIA

APRIL 2023

To my Parents, Dr. Aftab. M. Hussain for being a constant guardian and my dreams.

International Institute of Information Technology

Hyderabad, India

# Certificate

It is certified that the work contained in this thesis, titled "Labour Monitoring in pregnant women using Fetal Phonocardiography, Electrocardiography and Electromyography Technique" by Anushka Tiwari, has been carried out under my supervision and is not submitted elsewhere for a degree.

10th July 2022                                                    Adviser: Dr. Aftab M. Hussain

**Abstract**

Continuous monitoring of fetal and maternal vital signs, particularly during labor, can be critical for the child and mother's health. We present a novel wearable electronic system that measures, in real-time, maternal heart rate using phonocardiography (PCG) and Electrocardiography (ECG). Uterine contractions using electromyography (EMG).

When we monitor PCG for MHR and EMG for UC, using two different physical phenomena (sound and electrical impulses) reduces signal interference. The heart rate is determined by sampling for 6 seconds at 500Hz, using an autocorrelation algorithm, while uterine contractions are determined by sampling with the same rate and obtaining the duration and interval of contractions using thresholding. We calibrated the algorithms against known frequency sounds to determine an error correction factor. The resulting system can identify signal frequencies with an accuracy of ±5%.

When in later stages we employed ECG technique for maternal heart rate monitoring. The heart rate is determined using moving average filters to remove noises in the signal and ACF(Autocorrelation Function) for determining periodicity. For UC monitoring we stick to the same EMG technique. The resulting signal can identify heartbeats with an accuracy of ±2.9BPM and UC intensity within ±5 units compared to the gold standard CTC machine.

We also tried employing EMG technique to monitor the Fetal Heart Rate(FHR). But, in later stages of this design, this idea was aborted as we concluded that it needs further research on pregnancy stages and would require more intricate sensor integration that might not be in our reach at the moment.

The system is accurate, low-cost, and portable, so it can be deployed at primary healthcare centers in low-income countries. The system can also be used by women in the comfort of their homes. At the same time, the data collected is transferred to their doctor for analysis and diagnosis, which can bring a revolutionary change in the continuous monitoring of fetal well-being during labor.

Technological growth is achieving new peaks in today's era, especially in medical sciences. We will discuss the technologies used in health monitoring for pregnant women. Today, for fetal and maternal health monitoring, women have to go through long, tedious and not-so-healthy ways regarding regular checkups during pregnancy. These methods are either invasive or not restorative for long-term fetal health monitoring or costly enough that a middle-class woman cannot afford them. Even though good research is going on in this field to make the process easier, cheaper and safer, researchers have not achieved all three mentioned factors and good efficiency. Another major factor to consider is the portability of the system. Pregnancy is not easy for any woman, so going to the hospital to get a checkup often becomes challenging. It is about time to check on the baby in real-time while sitting comfortably at her home.

Index Terms—Labor monitoring, fetal heart rate, uterine contraction, phonocardiography, electromyography and electrocardiography.

## Acknowledgements

I thank my dissertation supervisor Dr. Aftab M. Hussain, for considering me in his group and helping me enhance my research qualities. Apart from being an accomplished trained supervisor from IIT – Roorkee, KAUST, and Harvard University, he is an amazing human being and friend and guide. He always gave me complete freedom to explore my strength, helped me overcome my weakness, and gave moral support to my family and me to pursue my further studies and achievements. I want to thank my friends Anis Fatema [currently a Ph.D. student at IIIT-Hyderabad, and Rishabh Bhooshan Mishra [Currently Project Assistant, Sensors and Actuators, Institute of Smart Systems Technologies, University of Klagenfurt(AAU)], Finally, I would like to thank my parents from the bottom of my heart for their limitless patience and support.

# Table of Contents

# List of Figures

# Chapter 1: Introduction

Technological growth is achieving new peaks in today's era, especially in medical sciences. We will discuss the technologies used in health monitoring for pregnant women. Today, for fetal and maternal health monitoring, women have to go through long, tedious and not-so-healthy ways regarding regular checkups during pregnancy. These methods are either invasive or not restorative for long-term fetal health monitoring or costly enough that a middle-class woman cannot afford them. Even though good research is going on in this field to make the process easier, cheaper and safer, researchers have not achieved all three mentioned factors and good efficiency. Another major factor to consider is the portability of the system. Pregnancy is not easy for any woman, so going to the hospital to get a checkup often becomes challenging. It is about time to check on the baby in real-time while sitting comfortably at her home.

The earliest method employed to estimate the FHR was the fetal-stethoscope which further developed into the fetoscope during the early 1900s. There are two types of fetoscopy: internal (endoscopic fetoscopy) and external. Internal fetoscopy uses fibre optic cable inserted in the uterus either trans-abdominally or trans-cervically for multiple purposes. While external fetoscopy employs a stethoscope to listen to fetal heart sounds. FHR was recorded using abdominal and intravaginal ECG (FECG) leads and was primarily used to assess fetal life for another half a century. In 1965, a continuous FECG monitor was proposed by Edward H. Hon, who measured the time intervals between successive R waves in the recorded heartbeat signal to calculate the heart rate[24]. Doppler ultrasound was developed a few years later, but its side effects due to continuous monitoring are undetermined. We will overlook several employed fetal and maternal health monitoring techniques in this section and discuss them in detail in the next section.

For over a century the domain of electrocardiography has been in existence. Still, even though there are remarkable advancements in adult scientific ECG, digital processors and digital signal processing techniques the study of fetal electrocardiograms is far from saturation [23]. Partly, this is due to lack of gold-standard repositories and due to relatively low SNR ratio of fetal electrocardiograms in comparison to the maternal electrocardiograms. This is impacted by various factors like the measuring electrodes, fetal heart along with the fact that fetal heart is weaker in comparison of the maternal heart resulting in lower signal levels. Also partly because of inadequate scientific understanding about fetal development and its cardiac function[25].

PPG (photoplethysmography) is a low-cost and simple optical technology that is able to detect the change in the volume of blood in microvascular bed of tissues [8]. It is employed non-invasively onto the skin's surface to make estimations. The photoplethysmographic signals are made up of pulsating and alternating anatomical waveforms due to the synchronous cardiac changes with each heartbeat in the volume of blood. This waveform is then superposed on a 'DC' baseline which is gradually changing with smaller frequency components due to the factors like thermoregulation, nervous system activity and respiration [27]. An interest revival has been seen in this technology in the recent times, which is due to the demand in more portable technology which is straightforward and low cost for primary healthcare centers as well as clinical settings based on a community. Photoplethysmography is an optical and non-invasive method that is able determine changes in the volume of blood vessels and hence measure the heart rate based on the light absorption change in the tissues at a specific red or NIR wavelength [26].

Continuous or intermittent monitoring can be done by external cardiotocography. A couple of transducers are placed on top of the maternal abdomen (one on the top position of fetal heart to estimate heartbeats while the other over the uterine-fundus to measure how frequent the contractions are happening) are employed to estimate the heartbeat of the fetus as well as the uterine-contractions. Cardiotocograph (CTG) is a paper strip that a doppler ultrasound provides to record the information. CTG was invented as a monitoring tool for detecting fetal hypoxia to help professionals take precautionary and necessary actions if needed. The motivation behind intra-partum cardiotocography monitoring contains reducing perinatal mortality and CP [28].

Doppler ultrasound technique depends on generating an ultrasound beam ranging in the frequency of 1-2 megahertz. The beam is yielded by an ultrasound transducer fixed on top of the mother's abdomen, penetrates tissue for reaching internal body structures, and the US transducer receives the reflected echoes. Currently, the most used technology to capture the fetal heart rate by cardiotocography is build on doppler ultrasound, where the ultrasound transducer is attached onto the mother's abdomen for continuous monitoring [32]. There are some technical limitations, such as frequent periods of signal loss and determining the beat-to-beat error of the fetal heart rate. For high body-mass-index mothers the signal loss is severe. Multiple gestations, premature deliveries, and the analysis of fetal heart rate recordings becomes critical around the second stage of labour [33]. A commercially available fetal heart rate monitor identifies six cardiac events: ventricular and atrial contraction, mitral valve closing and opening, aortic valve closing and opening. Moreover, these events can rarely be obtained in doppler signal from same cardiac cycle. Most of all doppler recordings can reliably detect around four cardiac events [10]. Based on insonification angle and the positioning of transducer, the obtained doppler signals may significantly differ from same fetal heart. The fetus's rotation and movement over time might impact doppler signal content. The cardiac wall movement is the most important component of doppler signal [11], with ventricular contraction having about 30% of the magnitude as compared to atrial contraction [10,11].

FMCG(Fetal Magnetocardiography) is a non-invasive technique used for monitoring and recording the magnetic-field activity generated by electric-field activity of the heart. FMCG is a safe technique as it does not emit magnetic energy fields. FMCG is a weak signal of around 10-12 tesla, much smaller than environmental magnetic signals such as the earth's magnetic field [36]. In fetal magnetocardiography (MCG), the magnetic-field present on top of the abdomen of a pregnant woman is monitored to gather the information on the electrical activity of the fetal heart. The fetal MCG can replicate the typical features of an ECG like QRS-complex and P-wave [35]. Sensitive sensors are needed as the fetal MCG is a weak signal, accounting for disturbances, such as fluctuations in the earth's magnetic field and the other fields generated by electrical devices around [13]. The magnetic field sensor is called SQUID (Superconducting Quantum Interference Device). Since this sensor is susceptible, it needs to be cooled by a liquid helium medium (-269°C). The sensor is operated in a flux locked loop since the flux to voltage transfer is not linear. The system can be used as a zero-detector by introducing a feedback system that employs the measured change applied to a feedback coil next to the SQUID. In this case, the magnetic flux passing through the SQUID is kept at a constant level. A flux-transformer is used to couple the inflow of the magnetic flux near the mother's abdomen into the SQUID [37].

FPCG (Fetal Phonocardiography) signal is obtained from the maternal abdomen by employing a transducer microphone and was proposed around ten years ago to monitor fetal heart rate as a passive methodology [16]. This technology makes frequent and real-time monitoring of fetal heart rate feasible. This technique was able to estimate the heart sounds from an envelope of signal and determined the fetal heart rate via time difference in the occurrence of the sound emitted by the heart. Subsequently, the accuracy of fetal heart rate determination depends on heart rate estimation accuracy. The execution of fetal heart rate detection degrades considerably if wrong sounds (accounted by different muscle sounds) are detected, or heartbeat sounds are missed [39]. The filtering technique is usually in pre-processing stages because of multiple interferences and the low-energy of fetal-heart sound. Fetal heart sounds originate from blood flowing in the fetal heart. Since the sound made by cardiac impulses are repetitive, fetal heart rate estimation can be defined as monitoring repetitive frequencies of heart sounds in continuous cardiac processes [14]. In the referred paper [14], the estimation of FHR is done by estimating the cyclic frequency and time-varying assessment using a sliding window. Repetitive frequencies are characterized by cyclic frequencies in the signal processing theory. The cyclic frequency of fetal heart sound if they are repeating every 0.4 seconds is around 2.5 Hz.

# Chapter 2:  Literature Search and Analyzation

## 2.1 Methods used for FHR:

### 2.1.1. Fetal Electrocardiography (FECG):

Fetal electrocardiography (FECG) is a non invasive way of measuring and monitoring the electrical activity induced by the fetal heart using, electrodes attached on top of the mother's abdomen. FECG can be monitored during labour and/or during initial weeks of pregnancy and is used to detect abnormal fetal heart rates (FHR) or patterns. [1].

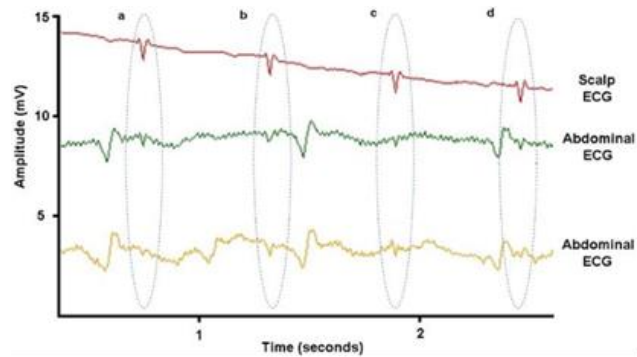

Figure 1: ECG recorded invasively (red line). Green trace: Electrode on top of the abdomen(closer to uterus). Yellow Trace: Electrode close to the mother's heart.

Fig. 1 depicts a segment of recorded FECG by both methods, invasively (upper trace) obtained by a fetal scalp electrode as well as non-invasively (lower two traces) by surface electrodes fixed on top of the maternal abdomen. Four fetal beats (labelled as a, b, c, and d) are highlighted (circled)on all the three traces. Because of the embedded broadband noise and the more significant amplitude artifacts due to the mother's heart, the abdominal traces contain much smaller fetal beats. Also, the artifacts are present in both, on top of fetal heartbeats as well as in between [1]. Due to the stage-wise activation of the myocardium, the ECG as monitored on the body surface results in PQRST-complex as represented in Fig. 2. This letter representation was first coined by Einthoven in 1895 [3, 4]. In this notation, the escalation of depolarization throughout the atria is represented by the P wave. During next 50ms, only very weak signals are recordable, as it takes some time for the depolarization front to travel through the AV node and since only a small number of myocardial cells participate in the atrioventricular conduction, the signals are very small [5]. Next, the QRS-complex is due to the depolarization of the ventricles. At the same time, the atria are repolarized; however, this repolarization is obscured by the depolarization of the ventricles. Finally, the repolarization of the ventricles takes place which accounts for the T wave. In some measurements, T-wave is followed by a small wave known as the U-wave, which for normal ECG, is believed to be due to the repolarization of the His-Purkinje system [4]. Other hypotheses for the origin of the U-wave have been suggested in [6]. The isoelectric segment of the ECG is defined from the end of the U wave till start of the next P wave, in which the myocardium does not have an accountable activity.

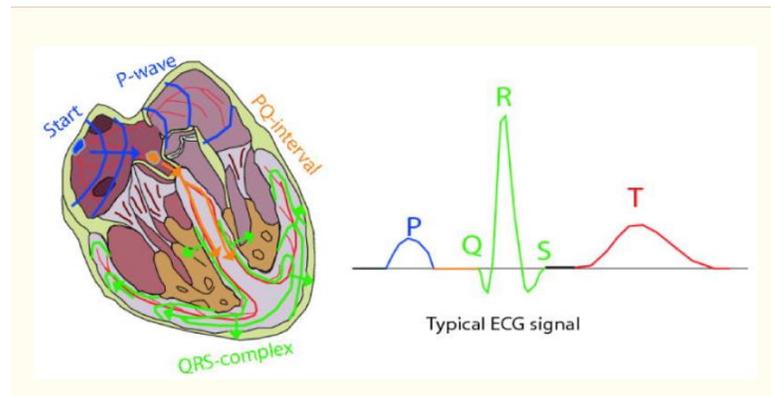

Figure 2: PQRST Complex waves explained [6]

Difficulties with FECG:

• For imaging specific defects, fetal electrocardiography hasn't proved to be a useful tool during labour. Howsoever this technique has been confined to issues such as general ischemia because of a certain fetal positioning that chokes the umbilical cord. [2]

• Signal interferences occurring due to maternal ECG signals (MECG, which usually has a magnitude of around 10 times higher than that of FECG [25].

• Artifacts caused by motion.

• Noises caused by muscles.

• Brain activity in fetus.

• Contact noise caused due to electrodes.

• Power line interferences along with other sources of noise. It is, therefore, necessary to have a signal processing system with a high signal-to-noise ratio.

## 2.1.2 Photoplethysmography (PPG):

Due to the broad availability of small semiconductor components at low cost and the advancement in computer-based pulse wave analysis techniques, a wide range of commercially available medical devices for measuring oxygen saturation uses PPG technology. Its also used in assessing autonomic function, blood pressure and cardiac output, along with detecting peripheral vascular disease [7].

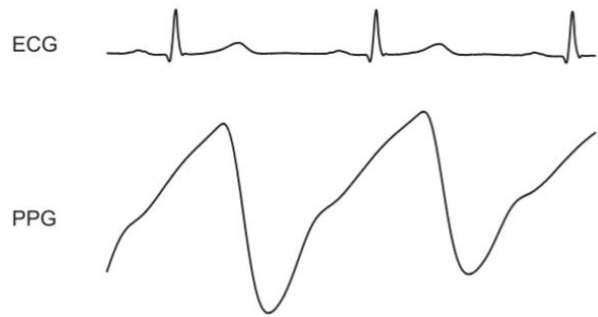

Figure 3: Signal waves compared to ECG and PPG [7]

This method can also be employed to measure the fetal heart rate through the maternal abdomen. Fetal heart rate is measured non-invasively by transmitting near-infrared (NIR) light of wavelength ranging between 650– 950 nm that can penetrate the human tissues with the peak penetration being at 890 nm via a photodetector to measure the reflected light. Modern PPG sensors employ low cost semiconductor technology with matched photodetector devices and LED's operating at the red or near-infrared wavelengths [8].

LEDs, essential source of light, convert electrical energy into light energy and have a single narrow bandwidth ranging around 50 nm. On average, the LED's intensity should be constant and preferably be sufficiently low to minimize excessive local tissue healing and reduce the risk of a non-ionizing radiation hazard. The choice of a photodetector is also essential [8]. The light source is chosen to match its spectral characteristics. Light energy is converted into an electrical current using a photodetector. They have fast response times compact, low-cost and are sensitive. Near-infrared devices can encase the daylight filters. The photodetector connects the low-noise electronic circuitry, including a trans-impedance amplifier and filtering circuitry. The size of the dominant DC component is reduced by a high pass filter and enables the pulsatile AC component to be boosted to a nominal 1 V peak-to-peak level [26].

Filtering circuitry is carefully chosen and is also needed to remove the unwanted higher frequency noise such as electrical pick up from (50 Hz) mains electricity frequency interference. It employs a trans-impedance amplifier design. Followed by the signal conditioning stages surrounding this, including low as well as high pass filtering and further signal inversion, amplification, and interfacing of signals. The design compromise here is the choice of high pass cut-off frequency which if often very crucial; the pulse shape can be distorted by excessive filtering, however AC pulse being dominated by quasi-DC component can happen due to too little filtering [7].

Another experiment performed by Zhedia and Bengb [7], which used the PPG technique for monitoring fetal and maternal heart rate, employed an approach based on adaptive noise cancellation (ANC). ANC uses photoplethysmographic signals extraction from the maternal abdomen for monitoring fetal heart rate.

Difficulties with PPG:

• This technique uses high power, generates heat, is expensive and difficult to implement.

• Another concern is the dependency of the signal quality on the source-detector (S-D) separation (optical shunt problem), which in turn depends on the type of source and photodetector.

• Fetal depth determines the signal quality from the surface of the abdomen. When the depth of the fetus of greater than 2.5 cm, the acquired signal has low SNR [27].

• The signal quality is strictly dependent on the position of the fetus in the uterus along with the position of the probe.

• The presence of motion artifacts and muscle contractions affects signal accuracy.

### 2.1.3 Cardiotocography (CTG):

Continuous or intermittent monitoring can be done by external cardiotocography. A couple of transducers are placed on top of the maternal abdomen (one on the top position of fetal heart to estimate heartbeats while the other over the uterine-fundus to measure how frequent the contractions are happening) are employed to estimate the heartbeat of the fetus as well as the uterine-contractions. Cardiotocograph (CTG) is a paper strip that a doppler ultrasound provides to record the information. CTG was invented as a monitoring tool for detecting fetal hypoxia to help professionals take precautionary and necessary actions if needed. The motivation behind intra-partum cardiotocography monitoring contains reducing perinatal mortality and CP [28].

Caesarean section rates are continuously increasing since past twenty years, which approves that CTG monitoring is required for the increment in caesarean sections. Rising cesarian sections in well developed countries contribute to rise in maternal morbidity and mortality [9]. When compared to a period before the invention of CTG to a period after the invention, it is seen that CTG monitoring helped women at risk more than any other methods of labour monitoring statistically [29]. A cross-sectional design was used for the remaining studies. Two high-risk groups were compared by Philpott and Stewart's research, but the two groups differed in their risk factors [30]. The results from two maximum-risk subgroups are combined. The study shows variation in monitoring types emerged because of the limitation in the number of monitors used.

Difficulties with CTG:

• Instead of contraction strength, external tocometry helps to show entire cycle of contractions along with the frequency.

• The absolute pressure value of externally placed tocometer are subject to the location and are not effective in people with higher BMI index [9].

• Internal Cardiotocography is a technique that directly connects fetal scalp to the electronic transducer. The wired electrode is connected to the monitor, going through the cervical opening with the other end connected to the fetal scalp . It is a dangerous technique as it may cause injuries in the fetal scalp and/or uterine infection [31].

### 2.1.4 Doppler Ultrasound:

For recording time less than about 20%, the fetal heart rate signal gets lost. Doppler signal analysis according to some reviews is described as state-of-the-art method for focusing on feature extraction and clinical interpretation. Due to just few available opportunities, limitations and challenges, the doppler ultrasound signal acquisition is not definable. We will look at the technical architecture of a doppler ultrasound-based fetal heart rate surveillance system and their connection to viable errors which might avert successful fetal heart rate monitoring [10]. Fig.4 illustrates these technical architectural blocks.

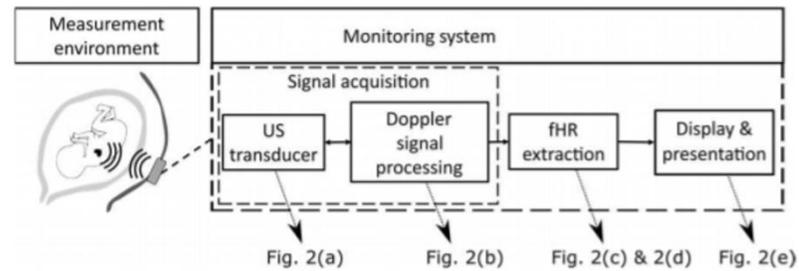

Figure 4: Technical building blocks for Doppler ultrasound [10]

The transmitted ultrasound waves propagate through maternal abdomen, based on the position of ultrasound transducer, interacts with tissue structure [32]. The reflection and transmission of ultrasound waves at the bounds of this pair of media with multiple acoustic attributes can be defined by the contrast in their acoustic impedance given by $Z = \rho c$, where $\rho$ and c are density of medium and propagation velocity of ultrasound, respectively. Ultrasound scattering in all different directions happens when ultrasound signals interact with defined structures, such as the red blood cells, which are lesser than their wavelength [λ=c/f0 with transmission frequency f0]. Disintegrated ultrasound signals are usually lower in magnitude in contrast to the specular reflections [34].

For fetal heart rate monitoring, primarily, the detection of doppler signal envelop happens and then applied to the fetal heart rate detection algorithm. The Hilbert-transform provides the analytic motion for envelope detection. The detected magnitude will be low pass filtered; however, different envelope detection methods are considered [33]. In conventional fetal heart rate monitors, the fetal heart rate was monitored from the time measured between peaks found in the enveloped signal of two different consecutive heartbeats. Moreover, due to multiple peaks and noisy signal in one cardiac cycle, the accuracy and reliability of monitored fetal heart rate were unsatisfactory [11]. The futuristic generation fetal heart rate monitoring device provides the solution by autocorrelation function for fetal heart rate estimation [11], [12].

Difficulties with Doppler ultrasound:

• The significant limitations of using Doppler ultrasound are its sensitivity to movement and not suitable for continuous monitoring.

• It also gives the averaged HR and cannot give the beat-to-beat variability.

• Doppler ultrasound is generally not used in the first trimester and is not always reliable due to signal complexity and the effects of fetal and maternal breathing[34].

• Min gestational age- 20-40 weeks.

## 2.1.5 Fetal magnetocardiography (FMCG):

In fetal MCG, the magnetic field atop abdomen of pregnant woman is monitored for retrieving information on the electrical activity generated by the fetal heart. Though other techniques exist (ultrasound, fetal ECG), either the reliability is low, or they do not provide significant information about the electrophysiology of the fetal heart, or they have a considerably low resolution which makes it complicated to employ them for specific medical applications [35]. The QRS-complex does the depolarization of ventricles; the P-wave reflects the depolarization of the atria, and the repolarisation of the ventricles is reflected by the T waves. The cardiac muscle cells give rise to a magnetic field and the volume currents flowing within the fetal MCG. The synchronic activity of many heart cells is measured, resulting in fetal MCG [36]. The magnetic field activity of a single cell is too bleak to be measurable. However, the overall magnetic fields generated by the heart cells are still weak (less than a millionth of the magnetic field generated by the earth). Immense calibration is required to reduce disturbances accounted by the magnetic fields caused by electrical devices and power lines and the earth's magnetic field fluctuations.[12]

The flux transformer is constructed using two superconducting coils that are wound but in opposite directions at a certain distance. The net amount of flux coupled into the flux transformer is null if this flux transformer is subjected to a homogeneous magnetic field. Therefore, flux-transformer is critically reactive for close sources with respect to the pickup coil and not as liable for sources located far away. The measurements are taken in a magnetically shielded room to avoid environmental noise [38]. The room that is magnetically shielded is made of an aluminium layer and another two layers of Mu-metal. The mother is asked to lay in a prostate position underneath the repository filled with liquid helium (cryostat) containing the SQUIDs and flux transformers to be close to the fetal heart. The monitored signals are given to a computational system which is placed outside of shielded room, where the signal processing occurs [37]. The maternal MCG interferes with fetal MCG as the signals are similar and are subtracted from the latter after data acquisition is made.

Difficulties using FMCG:

• It needs to minimize subject movement.

• Huge Equipment is involved.

• High cost required.

• Complex system design.

• min gestational age- 20-40 weeks.

## 2.1.6 Fetal Phonocardiography (FPCG)/Phonocardiography (PCG):

FPCG is an optimistic fetal heart rate monitoring technique. But the accuracy in the detection/monitoring rate of heartbeat burst sounds from an FPCG recording, have to be highly time-sensitive [41]. FHR is monitored using repetition frequency of the heartbeats detected, that is extrapolated without sound burst detection and filtering rather than from the cyclic-frequency spectrum peaks. Even if the SNR is as less than -26.7dB the repetition feature remains. If the SNR is lesser than -15dB this method clearly outperforms the previous methods [40].

The basic principle behind the working of FPCG is that the heart's mechanical activity is followed by the generation of a variety of characteristically essential sounds. These sounds are due to the opening and closing of heart valves and changes in the speed of blood flow that could provide diagnostic information. Sound transducers are placed atop the mother's abdomen, and hence in modern FPCG, FHS(Fetal Heart Sound) is picked through. Different characteristics such as frequency rate and changes in individual parts of the recorded cardiac acoustic signal can be measured.[16]

FHR estimation from FPCG analysis is difficult in extracting information from very noisy transducer data, and hence it becomes one of the main challenges faced. Data is also affected by acoustic damping, mainly due to the digestive activity and amniotic fluid and four primary sources of sounds such as maternal and fetal heart contractions, fetal movements and maternal breathing. Transducers' conditioning accounts for other secondary noise sources, including shear noises [39]. FHR becomes very challenging due to all these noises embedded in the FPCG signal in the time- and frequency domain that makes the extraction of FPCG signals from the sound transducers as low energy making the SNR is considerably low [15]. Various signal processing techniques and tools have been proposed and applied to extract valid information from FPCG recordings to address the problem mentioned above. FIR/IIR Filtering was used initially, which has low computational complexity yet a high chance of failure in separating the desired FPCG components in the entire signal. Some heuristic methods, such as spectral subtraction, enhance noise by employing the artefact attenuation implemented at low computational complexity [39]. Yet, they are mainly helpful in post-processing the FPCG signal. Adaptive filtering displayed poor performance in FPCG extraction. This could be refined if the maternal heart sounds (MHS) were first estimated and cancelled using a multi-channel system. Kalman filtering needs accurate reference signals, as it has high computational complexity [16].

Difficulties with FPCG:

• The most credible flaw of the FPCG design is the intense amount of filtering calculations and signal processing mathematics involved in processing the signal to give a clean fetal and maternal heart rate.

## 2.2 Maternal Heart Rate monitoring:

The techniques mentioned earlier can undoubtedly monitor both fatal and maternal heart rates. In fact, at a time, both are measured at the same time during practical use. Their processing techniques differ in how well that method can differentiate fetal-maternal and fetal heart rates. This part of the processing matters a lot and, if not paid attention to, can ultimately make the situation of life and death [20]. Ultrasound technology quite accurately calculates the FHR. The contrast between the two cannot be detected by the fetal monitor. Many times counting single heartbeat once, it will be counted twice, giving a completely wrong analysis of the data [43]. The confirmation fetal life is dependent on the admin of the fetal monitor, before monitoring the heart rate and verifying that the signal is coming from the fetus. It's straightforward for confirmation of source of the signal as maternal or fetal by contrasting audible fetal monitor sound with the maternal pulse [42].The chances of error are more when the fetal ultrasound transducer or a hand held doppler machinery is used, this is significantly in a case where the maternal heartbeat is ranging around 100BPM. Hence, simultaneous palpation should be used to reduce the chances of misidentification (using one's hands to check the body, especially while perceiving/diagnosing a disease or illness) for at least one minute of maternal radial pulse while listening to the device sounds [15]. This work, we present a robust data collection and analysis method for monitoring MHR and UC for women in labour. The filtering and analysis of the signal are performed in real-time using highly optimized algorithms so that they can be deployed on edge using simple microcontrollers. We employ ECG sensors to measure the maternal heart rate and electromyography (EMG) sensors for uterine contraction (UC), which is based on the electrical signals generated by the uterine muscles.

## 2.3. Technology to Differentiate the MHR From the FHR:

Three techniques are available to differentiate MHR from FHR. First, a $2^{nd}$ US transducer is fixed on top of the maternal heart. Mostly, MHR is estimated from the pulse oximeter on the EFM while oximeter probe is mounted on mother's finger [42]. Occasionally, to create a maternal heartbeat data, maternal electrocardiogram (ECG) device is utilised for the fetal monitoring [43].

## 2.4. Uterine Contraction (UC) Monitoring techniques:

### 2.4.1 Overview of Uterine Contractions:

Uterine activity may be assumed to be sufficient if the mother is already in labour, which is defined by continuous and progressive cervical descent and dilation, is occurring. Inadequate uterine contractions can lead to failure in labour progress. Excessive uterine activity may cause inadequate placental perfusion like in abruptio placentae. Furthermore, this gives rise to acidosis and fetal hypoxia. When it is necessary to augment or induce labour, the technician must be aware of uterine activity intensity, as overstimulation could lead to uterine rupture or even fatal compromise [44].

Manual palpation has been one of the traditional methods of monitoring uterine contractions in labour for many generations. This method can be employed to identify contraction duration and frequency, but it can only measure the intensity relatively. It requires constant evaluation, is time-consuming, and provides no distinguished record[45]. The effort can be highly tedious, and intermittent manual palpation occurs in most cases for short intervals. The contraction monitor significantly improves the accurate monitoring of the UCs. [18]. Contractions can be recounted by duration, strength, frequency, shape and uniformity.

**2.4.2 Non-invasive methods of uterine contraction monitoring**

• CTG – Cardiotocography

• TOCO – Tocodynamometer

• EMG – Electromyography

**2.4.3 CTG**

The fetal heart rate and uterine muscle activity are detected by two transducers placed on the mother's abdomen (one above the fetal heart and the other on top of the uterus for measuring contraction frequency). The information collected from doppler ultrasound records the cardiograph on paper-strip in real time. CTG cannot monitor the strength of the contractions. Here an integrated CTG uses a pressure-sensitive contraction transducer called a tocodynamometer while another transducer measures the fetal heart rate. The chance of getting external noise or mixed-up signals in this system is relatively high [46].

**2.4.4 TOCO (Tocodynamometer)**

The primary method of measuring uterine contractions is the tocodynamometer. Women in labour have traditionally been monitored by the tocodynamometer (TOCO). TOCO works on the pressure force produced by the contorting abdomen during a uterine contraction. The contractions are measured by a pressure change detecting sensor placed on the mother's abdomen. A belt is wrapped around the mother's abdomen for this arrangement, which can be very uncomfortable and constraining. The device can also be inaccurate, and its measurements could be subjective and dependent upon how tight the belt is wrapped. The intrauterine pressure catheter (IUPC), which is regarded as the reference (gold) standard, can be employed only after amniotomy, which means ruptured membranes [47].

### 2.4.5 EMG

In our opinion, the best way to measure uterine contractions is through EMG or Electromyography. In this method, the electrodes are placed on the abdomen, and they measure the muscle contractions in the form of electrical bursts. The signal function is based on the electrical activity replicating the uterine muscle contraction and relaxation. This is similar to the electrocardiogram, which measures the electrical activity caused by cardiac muscle polarization and depolarization [48]. A series of spikes continuing for a particular duration together generate a contraction. According to the studies done as a comparison of the three mentioned methods, it is found that the spikes formed by the muscle contractions of the uterus in labour are the smoothest in EMG monitoring and hence give better results than the traditional methods [44].

## 2.5 Literature Search Summary

In our understanding of the above-described methods for heart rate monitoring, the most efficient is the PCG (Phonocardiography), accounting for its achievable portability, low cost and very plausible signal processing scope. We have also incorporated monitoring of MHR using ECG technique from the abdomen, which gave highly accurate results. EMG technique is also a plausible way for heartbeat monitoring since it works on the principle of muscle contraction and relaxation, which again is very similar to ECG.

Differentiating FHR from MHR is still not wholly achievable and one of the factors that we will have to give honest thought to. Sustainable techniques are no doubt, but nothing non-invasive that gives reasonably good, distinguished results.

The best way to measure uterine contractions is through EMG or Electromyography. In this method, electrodes are placed on the abdomen, and they measure the muscle contractions in the form of electrical bursts. According to the studies done as a comparison of the three mentioned methods, it is found that the spikes formed by the muscle contractions of the uterus in labour are the smoothest in EMG monitoring and hence give better results than the traditional methods.

# Chapter 3: Hardware Design Architecture

## 3.1 Initial Circuit Design:

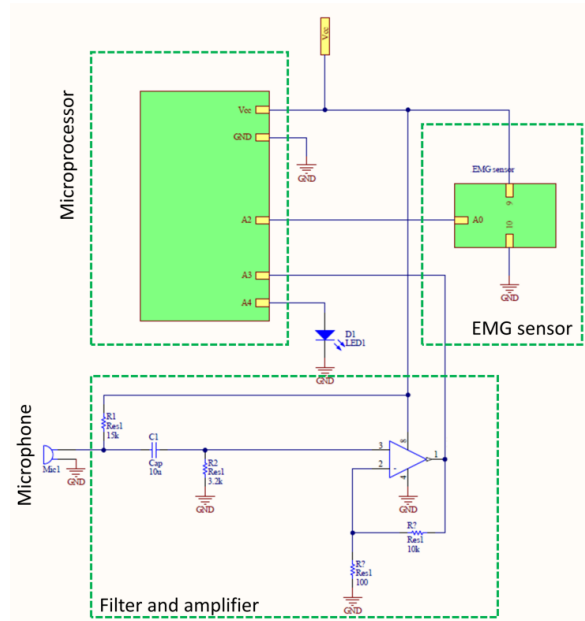

Figure 5: Circuit diagram for monitoring the UC as well as PCG.

Continuous monitoring of fetal and maternal vital signs, particularly during labour, can be critical for the child and mother's health. Generally, in medical grade integrated equipment for labour monitoring, oximeter is used for HB(Heartbeat). Our goal was to make an IOT integrated system that worked independently and wholistically on its own for labour monitoring. We present a novel wearable electronic system that measures, in real-time, heart rate using phonocardiography (PCG) and uterine contractions using electromyography (EMG). The above circuit design depicts the circuitry for UC and heartrate monitoring. We will incorporate MHR monitoring through ECG in later sections. Using two physical phenomena (sound and electrical impulses) reduces signal interference. The heart rate is determined by sampling for 6 seconds at 500Hz, using an autocorrelation algorithm, while uterine contractions are determined by sampling at the same rate and obtaining the duration and interval of contractions using thresholding. We calibrated the algorithms against known frequency sounds to define an error correction factor. The resulting system can identify signal frequencies with an accuracy of $\pm 5\%$. The system is accurate, low-cost, and portable, so it can be deployed at primary healthcare centers in low-income countries. The system can also be used by women in the comfort of their homes. At the same time, the data collected is transferred to their doctor for analysis and diagnosis, which can bring a revolutionary change in the continuous monitoring of fetal well-being during labour. There are several techniques for monitoring a child's and mother's progress and vital information during pregnancy. FECG, which relies on the electrical signals produced by the fetal heart, can detect aberrant fetal heart rates (FHR) patterns during pregnancy and labour [1, 2]. The cardiac movements are depicted as PQRST-wave complex, showing the cardiac muscles' depolarization and repolarization [3–5]. However, signal interference and noise caused by different factors such as maternal ECG (MECG, which is ten times more powerful than FECG), muscle activity noise in the mother, the activity of the fetal brain, electrode-to-electrode contact noise, etc., can introduce inaccuracies in the system [6]. Phonocardiography (PCG) uses heartbeat sound signals to determine heart rate. The sound signals emanating from the fetal heart are typically weak, requiring sophisticated algorithms to detect a periodic signal [13, 14]. This key disadvantage of the PCG technique results in the intense amount of filtering calculations and signal processing mathematics that is involved in processing the signal to give a clean fetal and maternal heart rate [6, 15]. We present a compact, wearable electronic system that undertakes most of the filtering and amplification processes in the analog domain before the digitization of the signals (Fig. 5). Thus, the signal fed to the microcontroller only requires simple computational algorithms such as autocorrelation to determine the frequency of a periodic signal.

PCG relies on the sound signals created by the beating of the heart. The signals were collected using a piezoelectric microphone. The analog signal produced by the microphone was filtered using an RC filter, which eliminates the DC level of the signal while maintaining the signal above the cut-off. We selected the cut-off frequency of 1 Hz because fetal heartbeat typically ranges from 1 to 3 Hz. After filtering, the signal was subjected to amplification using a positive feedback amplifier. The signal was digitized using a 12-bit ADC circuit with a reference voltage of 3.3 V. The data was collected in the form of a floating-point array. The samples were recorded for 6 seconds at a sampling rate of 500Hz, resulting in 3000 samples. The sampling was done for 6 seconds to ascertain that several heartbeat peaks were recorded in the sample so that that periodicity could be accurately determined. The system also consisted of an off-the-shelf, three-electrode EMG sensor which provided a single wire analog output proportional to the electrical activities detected. The sampling of the EMG signal was done with a sampling rate with a 12- bit ADC, also resulting in 3000 samples. The frequency and duration of uterine contraction can be determined from the received signals by running a simple thresholding algorithm on the samples in the array.

In the post advancement of the design which is mentioned explicitly in later sections. We incorporate MHR (Maternal Heart Rate) monitoring through PCG and ECG techniques while FHR (Fetal Heart Rate) and UC monitoring through EMG technique. Since the fetal heart signals are too bleak to be monitored by the PCG technique, it is efficient to capture the muscle activity of the fetal heart using the EMG sensors.

## 3.2 Final Prototype

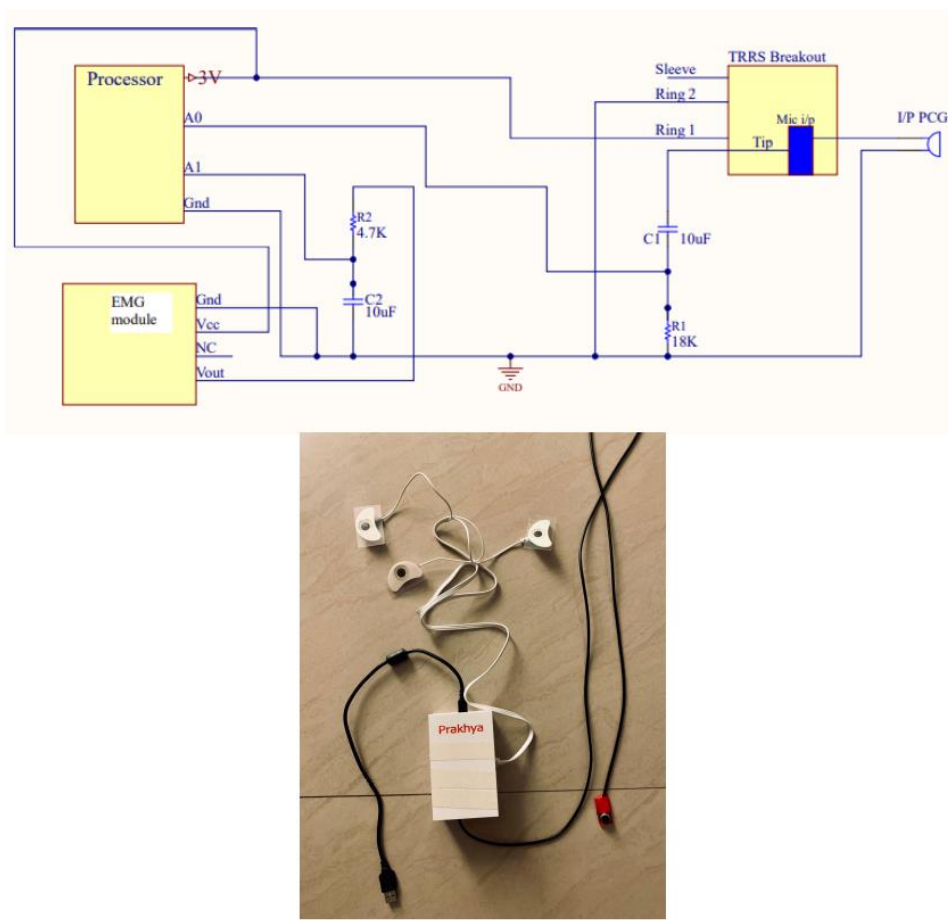

Figure 6: Simple overview of the final prototype circuit and the developed 3D printed prototype with all the sensors intact, ready for final hospital testing phases

## 3.3 Error correction

Because the algorithms for determining fetal heart rate and uterine contraction depend on the time difference between successive samples, accurately determining the sampling frequency is the most critical factor in such systems. By design, we were aiming to sample from the microphone at an interval of 1ms later increased to 2ms. However, if a delay of 1ms is introduced through the controller, the delay between two samples is invariably larger because of the time required to process the loop and the conversion time taken by the ADC. To account for the difference between the design and the actual sampling rate, we conducted the experiment for known frequencies ranging from 1 Hz to 3 Hz at an interval of 0.5 Hz. The obtained samples were subjected to auto-correlation using MATLAB to obtain the correlation plot, as shown in Fig. 6. Autocorrelation is a very robust algorithm for detecting periodic frequencies such as heartbeats [49]. The lag corresponding to the peak in the algorithm is the time period of the periodic signal, and hence its frequency can be computed [16]. Fig. 7 shows the plot for the experimental time period versus the ideal time period obtained for the signals at different frequencies. Because the line corresponding to the obtained data points was passing through the origin, it was clear that a fixed time interval gets added to the sampling time, irrespective of the signal frequency. This was expected because this time interval corresponds to the processing and signal conversion time. From the slope of the line in Figure 7, we determined the error in timing to be +0.86ms per lag step, which is why we started sampling at 2ms. Because this error was consistent, once determined for a specific system and controller, this error can be fed into the algorithm to obtain accurate values of the frequency of the periodic signal.

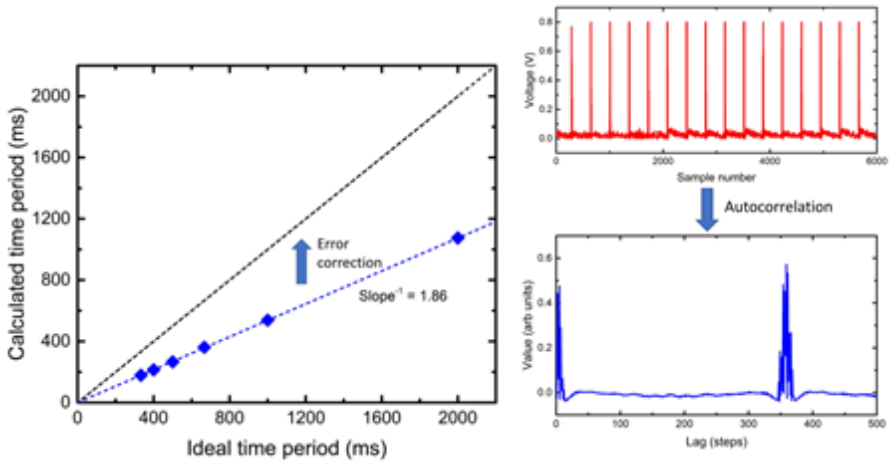

Figure 7: ideal time period vs error correction          Figure 8: sample plots vs their autocorrelation lags

## 3.4 Employing different filtering and amplification with a mic module

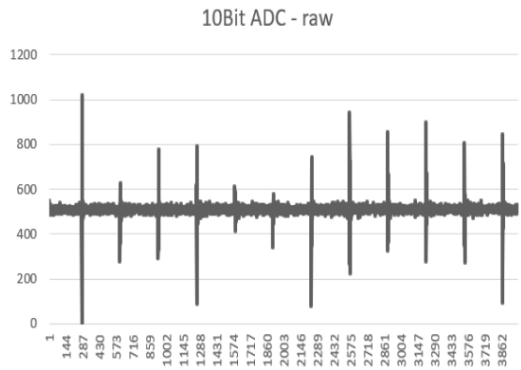

Figure 9: raw PCG signal on periodic monotones

Here, we haven't employed ADC to voltage conversion as in our experiment; we are concerned with the pitch of the signals rather than their amplitude. An observation in the shown graph (Figure 9) is that it's not starting with the origin; hence to compensate for that, we make the signal pass through a high pass filter to remove the DC. At this point, we are also thinking of removing the excessive noise that falls above the required set of frequencies keeping the human heartbeat frequency range in mind. The interesting thing to observe here is that the monotone is of frequency 100 BPM, and in the above-shown graph, autocorrelation can detect around 100.57BPM, which still stands accurate. Hence, we employ a bandpass filter with a frequency range 1Hz-3Hz. The following diagram is what we observe after putting a low pass filter of fH = 3Hz following a high pass filter of fL = 1Hz. Here we had a more particular result than expected, as shown in the graph below.

In the shown graph (Figure 10), we don't see any pattern whatsoever, unlike Figure 11, where we still see distinct peaks whenever monotone is played with the mentioned 100BPM frequency. With the same monotone here, we can see 89.94 BPM giving an error of more than 10 BPM. This observation opened new doors while researching burst sound frequencies and periodic signals.

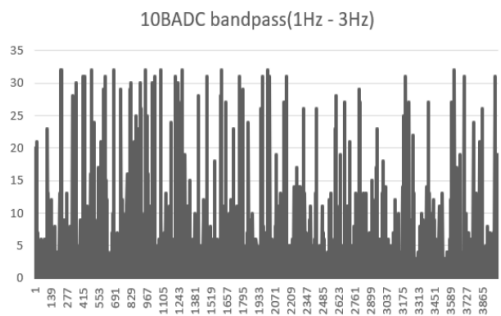

Figure 10: using a band pass filter with raw data

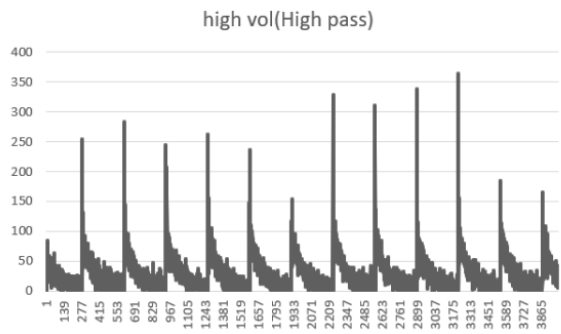
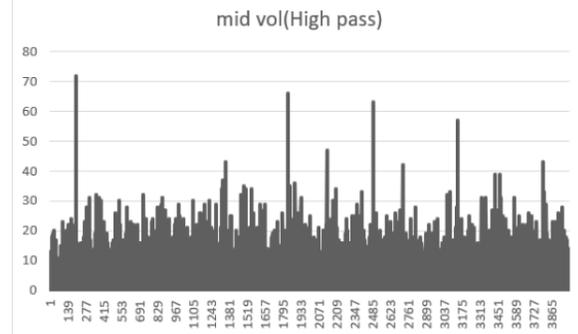

Figure 11: high volume monotone with high pass filter        Figure 12: mid-volume monotone with high pass filter

Even after employing and designing our bandpass filter according to our frequency bandwidth requirement, we aren't getting 100BPM precisely because of the frequency of the 'burst' monotone. The sound made by the phone's speaker, which 'replicates' the sound as designed by the application, is relatively high. This frequency ranges around 300-500Hz or more. This frequency was cut off as the low pass filter had a cut-off frequency of 3Hz. We then decided that the employment of a low pass after the high pass filter wasn't essential. After the above assertion, we found that even the human heartbeats are audible, which means they lie in a higher frequency range than 3Hz. The point to understand here is that we are talking about the 'sound' frequency of the heartbeat and not the periodic frequency of the heart, which is relatively lower. The further testing stages after this point were testing the setup with different volume ranges on the same monotone application again to understand the data better and know how our system reacts to varying pitches of the incoming sound signal. These results are depicted above in Figure 11 and Figure 12. As we see above, on mid-volume, a few beats of the monotone are still missed by the system, then we think of employing another amplifier stage and then a low pass stage to see what happens if we zoom in the signal and then clip off the noises. Below are the results we got after this point. These signals are depicted in Figure 13 and Figure 14.

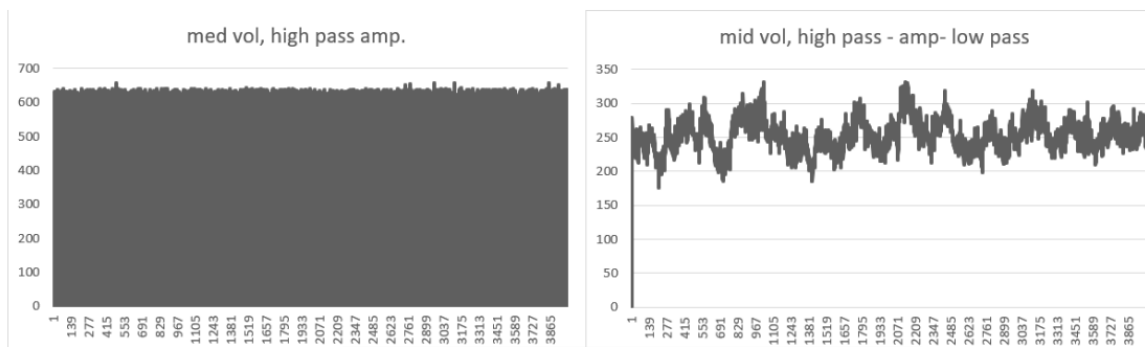

Figure 13: mid-volume monotone with high pass filter and amplification, Figure 14: mid-volume monotone with high pass filter and amplification, then a low pass filter

We tried to pull down the signal to the origin again using a high pass filter, giving us the signal as shown below in Figure 15. Here, we see that almost all the useful signal is being clipped off. Continuing the same input signal of 100 BPM frequency, the signal we got at this point was 122 BPM. Hence at this point, we change our approach further.

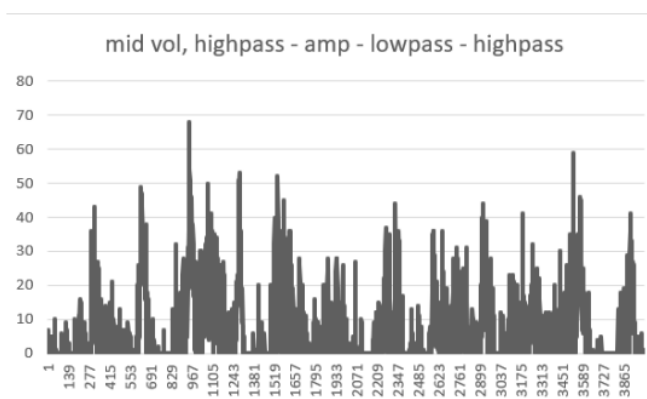

Figure 15: mid-volume monotone with high pass filter and amplification, then a low pass filter followed by a high pass filter

After this point, we move on to playing heartbeat audio sound from the phone to see the detection on our system rather than the previously used monotone of a defined frequency. At this point, we have made some very interesting observations. The data received from the heartbeat sound coming from the phone speakers made a very similar periodicity trait(visible beats on the graph for the same parameters) when seen graphically, as seen in the above graphs where the monotone is employed instead. Figure 16 depicts the same.

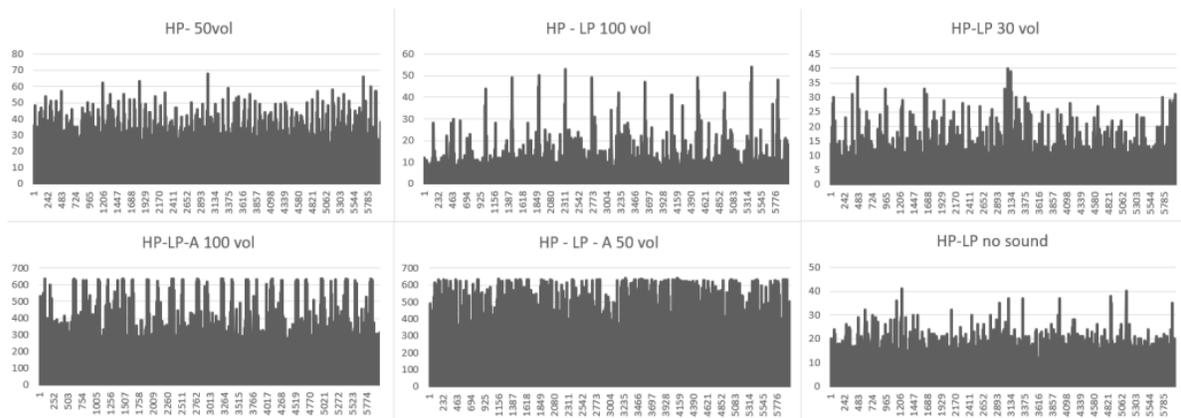

Figure 16: Data collected from the monotone with various monotone frequency

The periodicity trait discussed here is how well the system can capture periodic beats, as seen from the graph. We are taking these traits as the basis for choosing the following favourable path to enhance our data. We understood from the above data that the mic is capturing the sound that the phone's speakers already amplify. We won't see similar traits when we test our system with the heartbeat sound coming from the human chest. This is because the human body is an excellent conductor of sound. Sound travels almost four times faster in the human body than in the air, as sound travels faster in substances with relatively closer packed molecules. We might not need to create environments 'like' human heartbeat and should instead begin the actual human trials, and the mic should be able to capture it. The following section will see the results obtained by testing the system with actual human heartbeats. Before jumping onto the next section, the specific values of the high pass and the low pass filter as fL = 0.88Hz and fH = 482.28Hz, respectively.

## 3.5 True heartbeat detection through PCG

After the observations made in the last section regarding understanding the flow of sound in the human body, we concluded the human body provides a certain amount of amplification. As practical proof, we can listen to another person's heartbeats if we put our ear on their chest. Asserting that the sound emitted by the heart is amplified a certain number of times, making it audible enough for us to hear over another person's chest. With fL = 0.88Hz and fH = 482.28Hz as our constraints for high pass and low pass filter, we tested our system on our chest the following graphs (Figure 17) were the results.

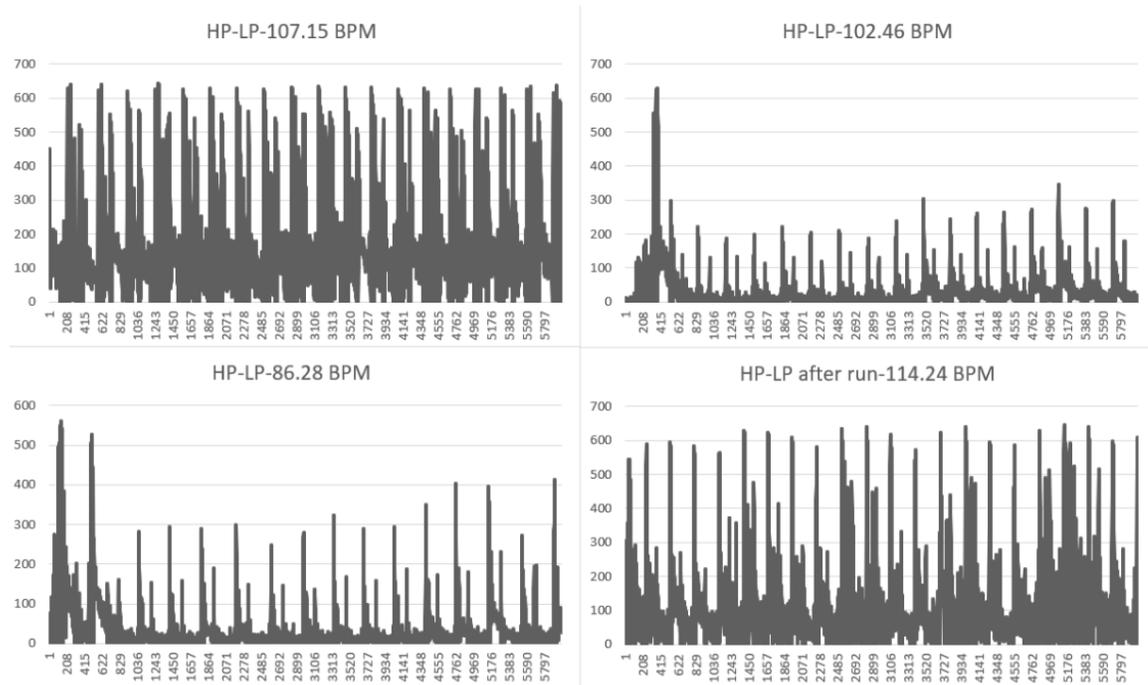

Figure 17: Frequency detected at different instances. The first image recognises the exact HB after excitement and after the different rest instances, as well as after a run when the prototype was stuck to the abdomen.

The average human heartbeat ranges from 60-120 BPM accounting for variation during activities like exercising, excitement, nervousness, etc. The results that we got here confirmed that PCG is a plausible alternative to Doppler ultrasound. The heartbeats calculated from the data and then auto-correlated to find the frequency came out as expected. This also proved that sound travels faster in the human body than air. The above assertion also logically fits as our body is 70% water. And water is a denser medium than air, making it more favourable for the sound to travel. We also tested our system with different songs to see if it could perform autocorrelation and give us any repetitive beats in the songs. Since we are recording our data in chunks of thousands and the period between each is around a ms, it is improbable to catch a repetitive beat in the small duration of 6s (6000 samples at this stage). The following testing step from this point was to check if our system could capture the human heartbeat if a song were playing in the background. The heartbeats that were measured at this stage came out to be as expected. The efficiency and robustness of our system match our requirement of the system being holistically robust. The results at this point are shown in Figure 18. We played a song at different times while the system was placed on the chest. Without the correlation, we can see the change in both the signals due to the background of the song, which holds numerous instruments playing periodically. In both cases, autocorrelation was still able to provide a plausible heartbeat BPM value.

The reason why we played a song in background rather than any other periodic frequency is because we wanted our focus to stay on extraction of periodic signal from a noisy background, as PCG employment would solely be done for maternal heartbeat monitoring. Any other further attribute monitoring would be done using sensors for muscle activity monitoring.

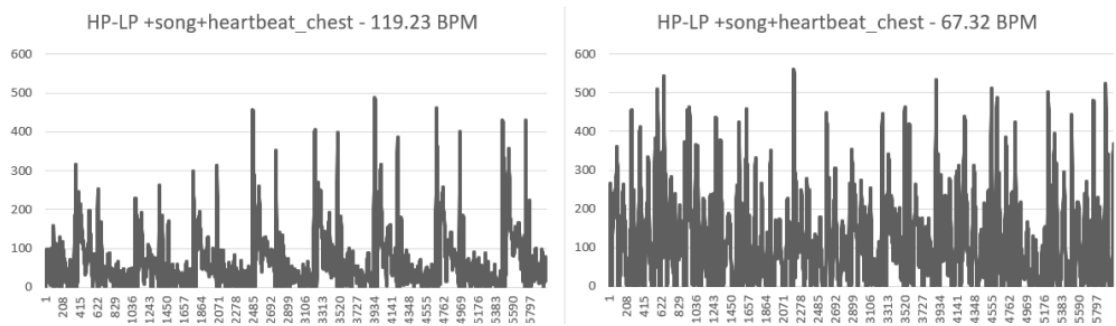

Figure 18: Heartbeat when the noise is introduced along with the sensor being attached to the heart position

## 3.6 PCB Designing

We will be discussing in detail how the device is working and understand how the flow of current is. This observation shall assist in understanding the working of the circuit along with having an intuition about the continuity too. Below is the PCB of our prototype. We will be discussing the functionality of each pin and how they are connected.

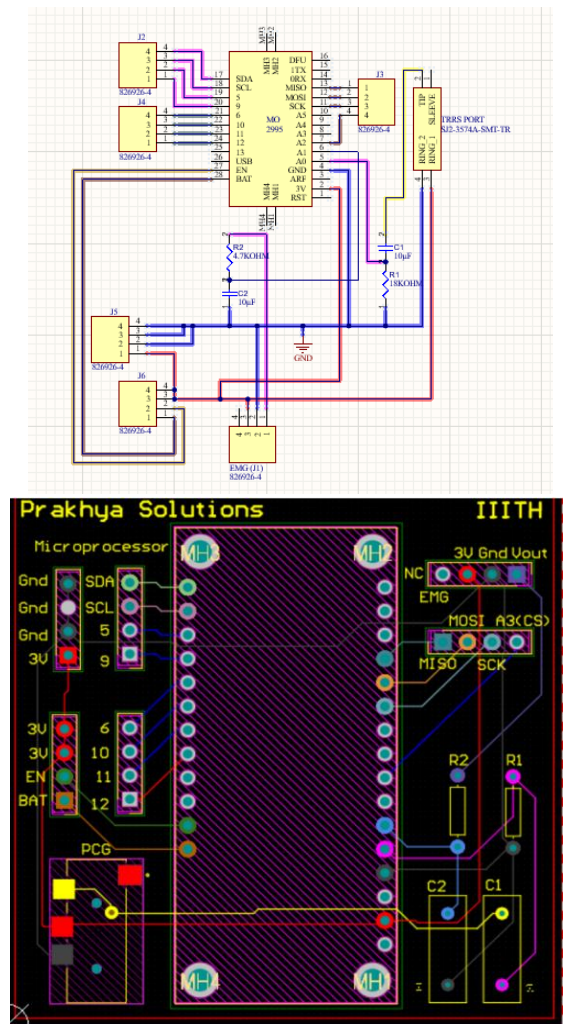

Figure 19: PCB designing on Altium software, using footprints for the processors, used sensors, PCG aux and for further SD card developments

The major connections are the ones connected to the analog pins of the processor and taking the sensor's data (PCG and EMG). We have made sure to colour-code our connections for better clarity. The yellow wire coming from the tip of the PCG sensor, which is connected to the electret microphone catching the signals, is connected to the filter stage consisting of R1 and C1. This connection then goes to A0 on our processor, which displays on the serial monitor and sends the data through bluetooth to our phones. Next, let's see the EMG connections. We have given a separate jumper set for the EMG signal coming from the EMG sensor. The NC pin is left free. Whereas the vout is coming from the output of the Grove EMG sensor that we are using. This signal is then connected to the filter stage consisting of R2 and C2, and then the signal is sent to the A1 pin on the processor. Capturing the PCG and EMG signals is the main functionality of our project. We have also given extra pins for future use of several communication protocols from the processor. We have SDA/SCL pins, Few GPIO's from the processor for using as CS pins or taking any additional signals. We have also given MOSI/MISO & SCK pins for any further developments using I2C communication. We have also made sure to have an ample number of extra GND and 3V pins for testing with LED's. Some major color codes of the wire are: Grey - Ground Yellow - PCG input Sky blue - EMG connection Red - 3V as shown in figure 19.

## 3.5 UC and MHR data collected from doppler ultrasound vs data from the prototype

The fabricated system is based on a microcontroller module with built-in analog to digital converters (ADCs), 32 kB SRAM and 48 MHz clock speed. The microcontroller's analog inputs were used to input signals from the ECG and EMG modules, and the signal was analyzed in real-time to determine MHR and UC. We used off-the-shelf modules for obtaining ECG and EMG signals using a three-electrode input.

Both the ECG and EMG modules provided an analog signal between 0 to 3.3 V that was input into the microcontroller using the ADC pins at 10-bit resolution. As shown in Fig. 20, the ECG electrodes were placed on the lower side of the abdomen, while the EMG electrodes were placed on the upper side of several subjects in late pregnancy (with consent and in accordance with Institute policy). To optimize the timing of the firmware, we separated the signal reading from the analysis part of the code. The timing of the signal reading part was controlled using a timer-based interrupt callback function, set to fire every 2 ms (500 Hz sampling). The signal capturing was done for 6 seconds (3000 samples), after which the timer was disabled to provide sufficient time for analysis of the signals. The two analog inputs provided 10-bit data stored in two arrays of 3000 samples of 16-bit unsigned integers (totalling 12 kB SRAM). This raw data was processed inside the controller to obtain the MHR and UC values. Thus, one value of MHR and UC was reported approximately every 7 seconds.

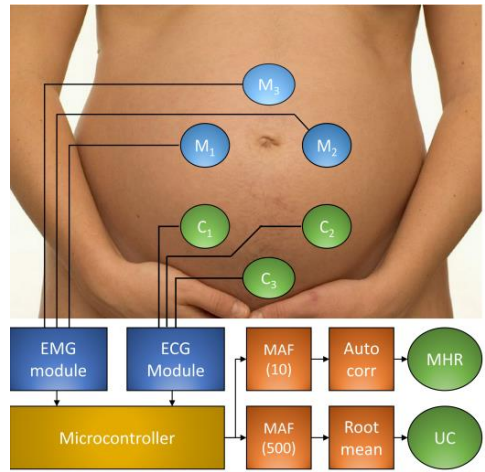

Figure 20: placement of ECG and EMG electrodes for monitoring MHR and UC, respectively

## 3.6 FHR data prediction

To measure the FHR, we have planned to take hospital permissions for our practical testing. For now, we tried to measure adult human heartbeats from the below-marked positions. The bpm of the subject measured via oximeter was found to be 90 BPM.

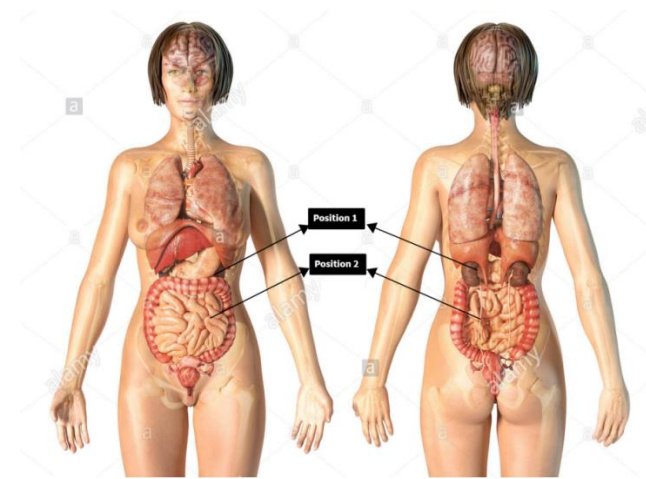

Figure 21: Data received with respect to the data received of the BPM monitored by the  PCG sensor attached to the defined positions.

The reference data is the heartbeat received by the oximeter(90BPM)

Below are the results we got:

For position 1: 87.24% accuracy.

For position 2: 78.73% accuracy.

## 3.7 Data captured from the device for FHR monitoring using EMG sensor vs CTG.

When captured with the PCG sensor, the data captured by our device for the fetal heart rate monitoring shows a high discrepancy due to the other sounds present in the maternal fetus when employed during the hospital testing on pregnant women.

We are already working towards employing an EMG sensor to monitor the fetal heartbeat instead of the PCG sensor and then employing the autocorrelation function to see the fetal heartbeats. Even now, when the data is looked at graphically, it is comparable with lower peaks.

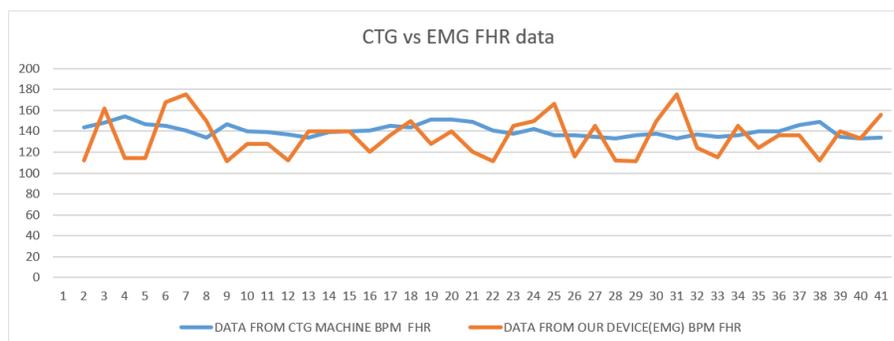

Figure 22: FHR data when compared to the CTG device

As the heart is a muscle, employing an EMG sensor instead of a PCG sensor is expected to provide the desirable results. But as of now, above figure compares the data received for a fetal heartbeat as compared to the doppler ultrasound (CTG). We are also trying to employ ML algorithms so that our device can predict and detect the higher frequency lower sound fetal heartbeats. The device is showing higher peaks yet comparable to the ultrasound, accounting for the various sounds made by the body of the mother apart from the heartbeat of the baby; hence once we try and employ EMG sensors instead of the PCG sensor, it is expected to show better and more accurate results. Even the result that we are getting now can be scaled down using filters to match the CTG output of the signal. In later stages of this design, this idea was aborted as we concluded that it needs further research on pregnancy stages and would require more intricate sensor integration that might not be in our reach at the moment.

# Chapter 4: Arduino(IOT) Design Architecture

To summarise the work done till now, we started with an in-depth literature search, which led us to decide on the favourite technique for our problem statement. We concluded that amongst other methods fitting our requirements, using PCG for heart-rate monitoring and EMG sensors for labour monitoring seemed the most feasible, keeping cost, portability, and robustness as considering factors. We then implemented our PCG circuit using a primary electret microphone, high pass filtering, and amplification stages to get the required frequency detection using the monotone frequencies from 0.5 to 3Hz. After successfully mentioning those above, we had the proof of our chosen technique. In the same stage of development, we also did the error compensation for our processor to see the period of our recorded heartbeat data accurately. This error compensation took care of the fraction of a ms taken extra by the processor to process the number of data points dumped in it in one instance of the cycle. After this point, we shifted from using a stand-alone electret mic to a mic module with an electret microphone and an amplification module attached to a PCB. The amplification module makes the amplification of the initial input signal dynamic. In the following sections, we will look into the real scenario implementation of our system.

## 4.1 Stage of code development

Till now, we have seen the hardware tweak and changes that led us to detect human heartbeats. We have been integrating MATLAB and Arduino for this result. The heartbeats are captured serially through the processor to the Arduino and the autocorrelation in MATLAB to find out the exact time period of the heartbeats. There were three major advancement stages in the code of our system. The first one where we started was monitoring the heartbeats and autocorrelating them to find the required time period. The second was introducing the adafruit timer in our code design and the third was making the code feasible of further SD card implementations.

Figure 23: Depicts the code snippet with the defined variables as well as autocorrelation algorithm defined.

Rawdata_HB and rawdata_UC are the variables used to capture the heartbeat and the UC signal from the processor, respectively. The autocorr[lags] gives the autocorrelation for several defined 'lags.' Mini and maxi are used for averaging and shifting the signal dynamically throughout the loop, as seen in the above code snippet. Autocorrelation is one of the most robust algorithms for finding the time period of periodic signals. It shifts the main signal with a lag. Lag is the unit by which the signal is shifted. The shifted signal is multiplied with their corresponding impulse, and the value is added. This is the i'th correlation point for i'th shift. The lag at which the autocorrelation function gives the highest peak is the point where the signal properly intercepted with the original signal and hence the point which will give us the time period off the signal when multiplied with the delay between each consecutive impulse. The following is how we implemented the same on our code. In this code, we have not yet accounted for uterine contraction code(UC), which will see in the next version of the code.

```
//Autocorrelation-----------------------
  for(i=0;i<lags;i++)
{
  for(j=0;j<n_HB;j++)
  {
    if(i+j<n_HB) autocorr[i]+=rawdata_HB[j]*rawdata_HB[j+i];
    else autocorr[i]+=rawdata_HB[j]*rawdata_HB[n_HB-j+i];
  }
}

//Peak detection-----------------------------
j=0;
peak=0;
peaklag=0;
for(i=180;i<lags;i++)
{
  Serial.println(autocorr[i]);
  sum+=autocorr[i];
  j++;
  if(autocorr[i]>peak)
  {peak=autocorr[i];
  peaklag=i;
  }
}
Serial.println();
bpm = 60000/(peaklag*ERR);
```

Figure 24: autocorrelation defined with respect to our defined variables as shown in the above snippet

In monitoring uterine contractions, we needed to find the duration of the contraction rather than monitoring if any of the peaks due to muscle contraction and expansion passed a certain threshold. What we implement here is if a full contraction or, in our situation, a muscle cramp exhibited a pitch higher than a certain threshold. We subtract the average in the code snippet below to pull the signal by that value to normalize and visualize better. In contrast, we are later squaring to remove all the negative impulses and see the signal above the origin, normalized at all times. Here the UC_club is a variable with a value of 10. We are checking the signal at the interval of 10ms ten times to make a signal 'a' to process further. UC is not a time-sensitive signal like heartbeats.

```
void Find_UC()
{
  aver = sum/n_UC;

  for(i=0;i<n_UC;i++)
  {
    rawdata_UC[i]=round(rawdata_UC[i]-aver);
    rawdata_UC[i]=rawdata_UC[i]*rawdata_UC[i];
  }

  for(i=UC_club;i<n_UC;i++)
  {
    a=0;
    for(j=0;j<UC_club;j++)
    {
      a+=rawdata_UC[i-j];
    }
    if(a>threshold)
    {
      if(flag==0)
      {
        UCstart=i;
        Serial.print("UC started = ");
        Serial.println(i);
```

Figure 25: Defines the smoothening of the signals received with respect to the range as expected the hospitals.

Next, the 'if' loop will keep checking until the signal is above the threshold. Once it falls below the threshold, we will mark that time as the duration of that contraction. We will wait for 30s to see if there is any other contraction. If not, we will state the last contraction as the latest and define the time at which the contraction occurred by the above-stated logic. The total time till which the loop waited for another contraction is given by t. If there is a contraction in the meantime, then that becomes our latest 'signal, and we monitor till what time that signal is above the threshold, repeating the same process over again. The same is shown in the below snippet.

```
if(i-UCend<min_durr)
{
  UCstart=UCstart_prev;
  k--;
}
else
{
UCstart=i;
Serial.print("UC started = ");
Serial.println(i);
}
}
}
}
else
{
  if(UCstart!=0)
  {
    UC[k]=i-UCstart;
    UCend=i;
    k++;
    UCstart_prev=UCstart;
    UCstart=0;
    Serial.print("UC ended = ");
    Serial.println(i);
```

Figure 26: UC monitoring based on the impulses based on the muscle contraction and relaxation based on the data received by the EMG sensors

At this stage, we are processing 2000 UC's and 4000 HB's in one loop. In the next stage, we will optimize the timing of the signal using an internal timer and interrupt our processor.

Figure 27: The timer analysis based on the monitoring and processing of data.

Adafruit_Zerotimer is a pre-set adafruit library used for programming the timer and the interrupt for Arduino ide. Here we can set the prescaler based on the frequency of the timer. On line number 122, we are giving a 'compare' value to our library, according to which it will time the signal. Prescaler and compare variables are fixed in the setup part of the code. On line 124, the setCallback function calls *callback_func pointer, which points to the function that needs to be called when the timer stops at the defined value. In simpler words, interrupt is raised when the program stops at the predefined timer time, which we specify in compare and Prescaler adjustments in the setup part of our code. The above snippets were from the library for understanding our Arduino code better.

```
tc_clock_prescaler prescaler = TC_CLOCK_PRESCALER_DIV64; //pre-scale by 64 -> 750 kHz
uint16_t compare = 750; //750 for 1 ms timer

// timer tester
Adafruit_ZeroTimer zerotimer = Adafruit_ZeroTimer(3);

void TC3_Handler() {
  Adafruit_ZeroTimer::timerHandler(3);
}

void timer_enable()
{
  zerotimer.enable(false);
  zerotimer.configure(prescaler,        // prescaler
        TC_COUNTER_SIZE_16BIT,       // bit width of timer/counter
        TC_WAVE_GENERATION_MATCH_PWM // frequency or PWM mode
        );

  zerotimer.setCompare(0, compare); //(channum, compare)
  zerotimer.setCallback(true, TC_CALLBACK_CC_CHANNEL0, timer0CB);// timer0CB is the event that triggers a callback
  zerotimer.enable(true);

}
```

Figure 28: 1ms timer settings.

The TC3_Handler takes care of the pin specifications of the processor accordingly. Here zerotimer is the instance we made for the library, and timer0CB is the event that is called when the timer stops. The timer enable is used every time we need to capture data. So whenever timer enable is called, the program will jump to timer enable, and execute it for the defined time (using Prescaler, compare etc.). When the timer stops, program will jump back to the original processing that it was doing. The function timer0CB is just capturing the data. The next advancement we make before configuring the Bluetooth is making 'n' common or the number of samples common instead of separating 2000 for UC and 4000 for HB.

```
//timer call back--------------------------------
void timer0CB()
{

  rawdata_HB[HBs] = analogRead(A0);
  if(rawdata_HB[HBs]<mini) mini=rawdata_HB[HBs];
  HBs++;

  rawdata_UC[UCs] = analogRead(A1);
  UCs++;
  UCstart++;

}
```

```
void BLEsend()
{
  for(i=min_lags;i<lags;i++)
  {
    ble.println(autocorr[i]);
  }
}
```

Figure 29: Introducing the timer settings to our autocorrelation, letting the processor know when to monitor the readings and when to process.

In the above code snippet, we see the bluetooth module of our processor for low-energy data transmission to our phone. BLE of adafruit is made user-friendly and easy to understand, as described in [50]. Whatever data we want to send should lie inside 'ble.println'. The only requirement of sending data through Bluetooth is that each ping should be sent in the next line. Hence, the use of println. By default, it sees the end of the line as the end of the data ping in one go. BLE (Bluetooth low energy) employs UART for transmitting the data serially to the mobile application. This is the primary development of our entire system.

## 4.2 Modifications of the above code

In the later stages of development, we made a few changes to the core code of our prototype. We decided to keep the total number of samples gathered as 3000. It implies 3000 samples for heartbeats and uterine contractions (UC) impulses. We have also defined the total number of lags as 550, for 54.5BPM (the minimum reference-range for the heartbeats). And we see the highest peak between lag 300(100BPM) to lag 167(180BPM) for monitoring the fetal heartbeat. Since we also have to keep smoothening the signals while monitoring, we choose to smoothen six samples at a time to disqualify the noises due to additional muscle impulses. These minor changes are set as predefined parameters to make our code more robust. We have also included the SD card integration using the SPI pin given with our processor and extended on to our PCB at pin 10.

After collecting the UC intensity, the averaged sum of the data collected on our analog pins connected to the EMG sensor. We now had a task to modify this data in the range of 0-120, which is accepted by the hospital equipment reading and more valuable to the doctors.

To do those mentioned above, we pass the UC intensity with the smoothening filter and manually correlate the data gained by the doppler ultrasound monitor closely. This gave the basis for estimating our values ranging from tens of thousands to millions in a range of 0-120. We shall see the results of the same in the later results section.

Everything else is closely related to the above explanation of autocorrelation and sending the data in the 'void loop' predefined function of the Arduino, which makes those particular functions inside the loop run endlessly.

Another development was to use the EMG sensor to receive the FHR, as the EMG sensors capture all the data made by the muscle activity in that particular region it covers. These sensors are placed near the mother's abdomen, which is close to the fetal heart. Whereas the PCG sensor robustly helped gather the data on the maternal heartbeats.

We then apply our lag conditions to our code, which give us the peak and hence the time period or the BPM of both mother and fetus using PCG technology for the prior and the EMG sensors for the latter.

Now that all our values are in the range as expected, we send BPM (mother), BPM (fetal) and UC intensity ranging from 0-120.

## 4.2 Receiving Data through Bluetooth

We have also implemented a few tweaks in the android studio application provided by the adafruit for receiving the data through Bluetooth so that we see the plot and the incoming data on the same page in the mobile app. Below is how the graph(Figure 30) looks post connecting the phone via Bluetooth to the processor and navigating to the UART button.

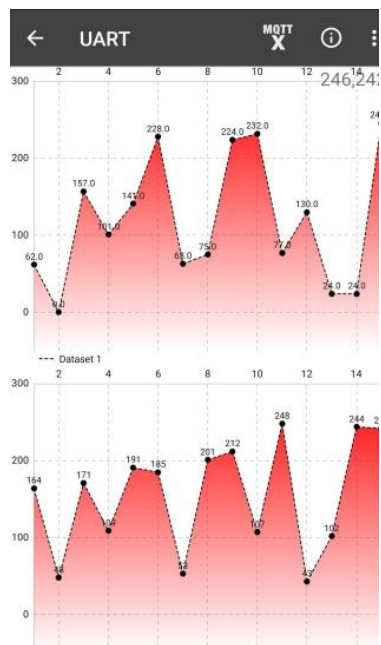

Figure 30: Application receiving the data as sent from the processor. Confirming the data received from the processor, is as expected for both UC as well as HB

This screenshot is the earliest development of the application to confirm the data that we send through Bluetooth is accurately received or not. In the depicted picture of the mobile application, on the top right corner is where we see two unsigned int values separated by a comma, according to our requirement. The top graph depicts the 'a', and the bottom graph depicts 'b' in the incoming data of the format (a, b) as shown in the top right corner. The reason why we are not doing any form of computing on the mobile's end is to keep the computing as fast as possible near memory, which will help us to achieve near real time data analysis more efficiently.

## 4.3 Verification of EMG data for employment in UC monitoring

We tried the EMG system on the forearms and abdomen to see how the muscle reacts when we contract and relax the muscles to activate them. For activating the forearm and the abdomen muscles, we place the electrodes as depicted below.

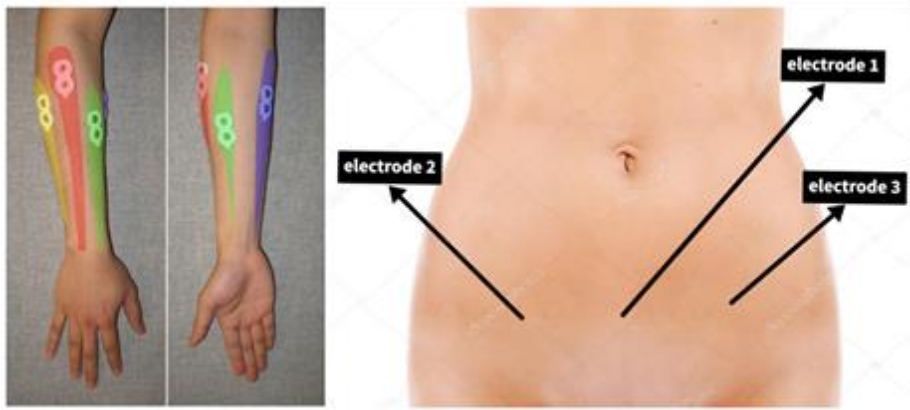

Figure 31: Defines the position of the stuck of electrodes for the UC monitoring (EMG electrodes stuck to the abdomen).

And below are the results we got from the forearm flexing:

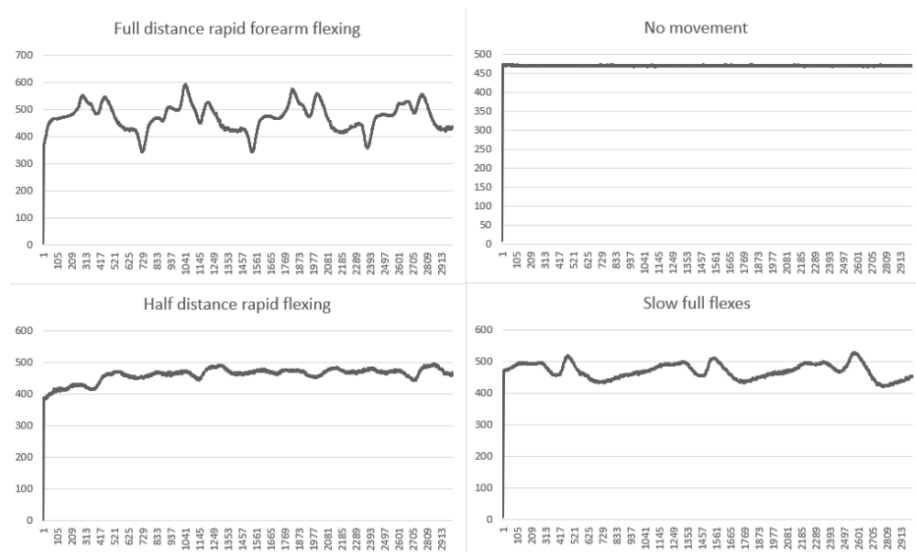

Figure 32: Data received as per the flexing the forearm flexing.

The above are the results we got with the movements defined in Figure 32. Full distance and half distance flex define the circumference of the arc covered by the arm from the elbow. We can clearly see the muscle movement from the graph depicted above. Next, we see how the abdomen muscles activate during certain motions.

At this stage, we also want to see the muscle activation of the abdomen to get an idea of how well our system is able to monitor the uterine contractions. The exercises that we will employ in this stage would be as follows:

1. Rapid contraction and relaxing of the abdomen moderated by breathing.

2. Slow contraction and relaxing the abdomen moderated by breathing.

3. Deep breathing.

4. No motion.

Below are the graphical results of performing the above exercises as depicted in figure 33.

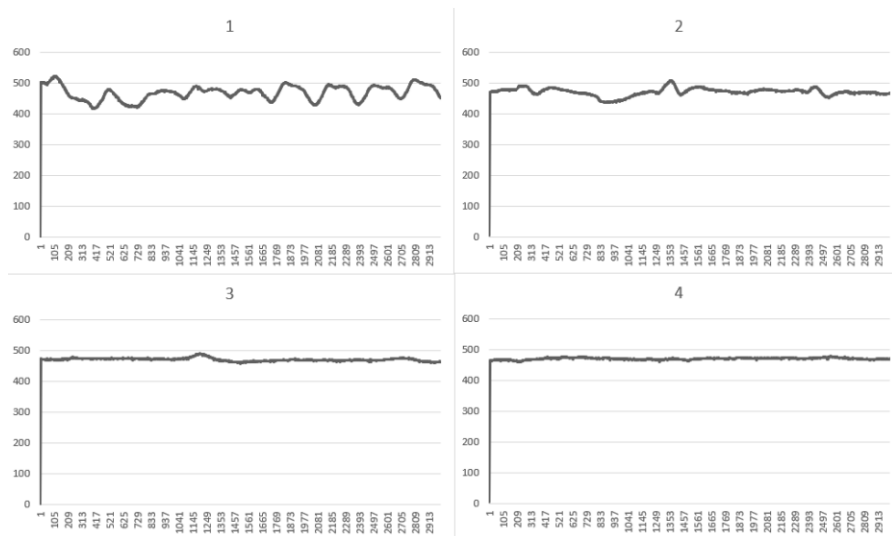

Figure 33: Data received when abdominal muscles are activated defined on the positions defined.

Delta($\Delta$) is the difference between the highest and the lowest peaks of the above graph. The respective delta values are:

$\Delta 1 = 81$

$\Delta 2 = 48$

$\Delta 3 = 33$

$\Delta 4 = 6$.

## 4.4 Algorithm for MHR monitoring

It was observed that the collected ECG data had high-frequency noise and significant baseline wander, which are known to be caused by electrical noise and from the variation in skin-electrode impedance [22]. We applied a moving average filter (MAF) of window width 10 on the raw data to remove high-frequency noise. This provides us with a very clean signal with a clear PQRST wave complex, as shown in Fig. 34(a). However, the baseline wander makes the data unsuited for applying an autocorrelation function to calculate heart rate. As the baseline wander is essentially just low-frequency noise, we applied another MAF of window size 280 to extract it and subtract the resulting array from our signal (Fig. 34b). This step is essentially a high pass filter, and the window is kept wide enough to ensure that none of the signals of interest is removed. This method for applying a bandpass filter is both computationally light and has low memory requirements making it ideal for a microcontroller implementation. After the removal of noise from the signal, an auto-correlation function was applied to determine the heart rate as it is immensely robust in detecting the periodicity of signals. We found the highest peak in the autocorrelation output in the 240 to 670 lags range, representing 125 bpm and 45 bpm, respectively (the range for human heart rate). The position of the peak lag Was converted into frequency using the sampling rate to obtain heart rate in bpm.

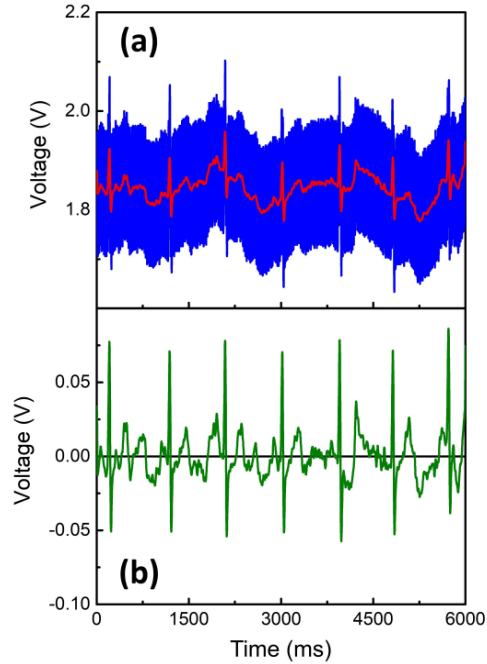

Figure 34: (a) Unfiltered (blue) and filtered (red) ECG data. (b) After baseline wander removal.

## 4.3 Algorithm for UC monitoring

EMG signals also suffer from baseline wonder while also having noise originating from skeletal muscle contractions. These were processed by first removing the low-frequency noise and DC offset with a wide window MAF of width 500 that is subtracted from the samples. This retains the F2 wave in the EMG signal. Because we do not retain the F1 wave, we cannot determine the UC intensity for low-intensity contractions. However, using the intensity and frequency of F2, we can determine UC intensity for strong contractions, which is needed for diagnostic purposes. As a measure of the intensity of F2 wave, the square root of the sum of the absolute values of the filtered signal was taken, which gave a strong correlation with CTG values[19]. This was scaled down by a factor of 9 to match the ranges of the TOCO device. Thus, UC intensity was calculated from the raw data (x) as:

$$UC_{int} = \frac{1}{9} \sqrt{\sum_{k=0}^{3000} \left| \frac{1}{N} \sum_{j=k}^{N+k-1} x_j \right|}$$

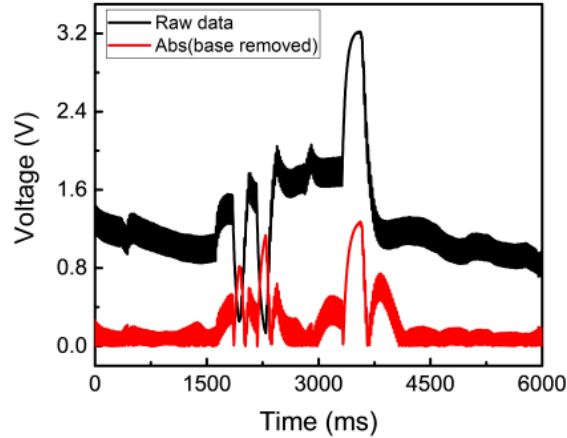

Figure 35: EMG signals obtained for a uterine contraction before (black) and after (red) the removal of baseline wander.

## 4.4 SRAM considerations

Techniques such as the Fast Fourier Transform (FFT) are memory intensive for such deployments, using 2×F F T length values even for an in-place implementation and needing float type values, which are at minimum 4 bytes. Because it would take 24 kB of memory for a 3000-point FFT, it would not be possible on a microcontroller with 32 kB SRAM without removing some of the peripherals. It was, therefore, avoided in this work. For heart rate calculation from ECG, we used 6000 bytes for the raw data, 580 bytes for the two filters, and 1340 bytes for the autocorrelation vector, which totals around 11 kB of RAM. For UC monitoring, we used 6000 bytes for the raw data and 1000 bytes for the MAF. Thus, the entire algorithm and all the peripheral functions can be run using around 25 kB SRAM.

# Chapter 5: Results

## 5.1 FPCG Results

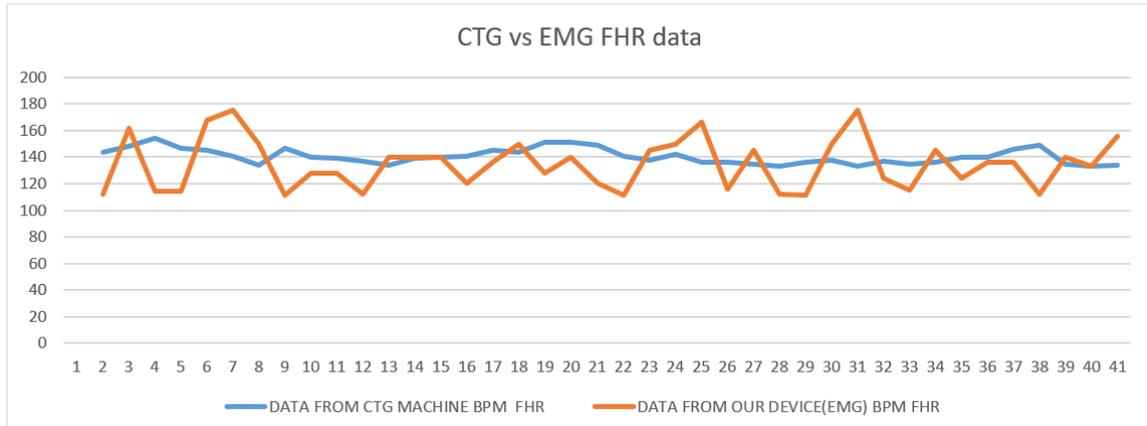

Figure 36: FHR using EMG(Orange) vs CTG(Blue)

We tried employing FHR monitoring using EMG sensor, but since a lot of muscles are at play during pregnancy, it was hard to achieve medical grade accuracy using EMG sensor for FHR monitoring.

## 5.2 MHR Monitoring using PCG

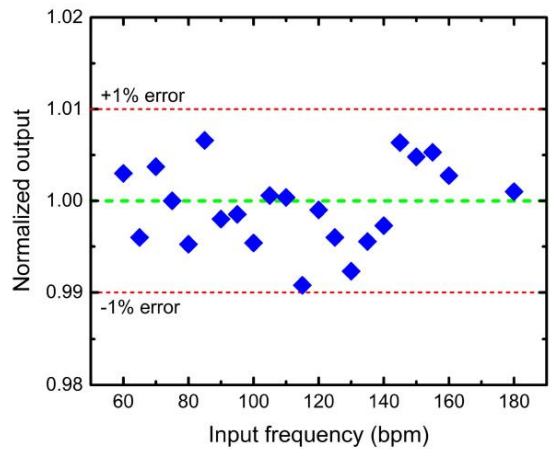

Figure 37: Error graph for MHR monitoring using PCG when the mic is close to the heart of the subject.

The normalized output (calculated frequency/input frequency) shows the accuracy with which the fabricated system is able to detect heartbeat signal frequency using PCG technique. ***The error is within 1%.***

## 5.3 MHR monitoring using ECG & UC Monitoring using EMG: Efficiency

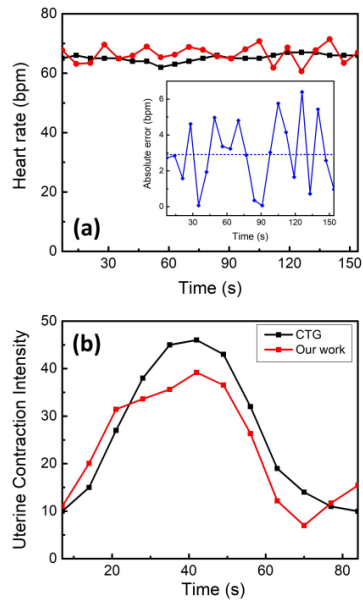

Figure 38. (a) Heart rate data for a female subject as calculated by our device (red) compared to a medical-grade oximeter (black). (b) Intensity of a UC as reported by our device compared to a CTG machine

As shown in Fig. 38a, our system detected the heartbeat of a female subject with an average absolute error of ±2.9 bpm when electrodes were kept at the mother's abdomen. The calculated values (red) are compared with a medical-grade oximeter (black). In the case of UC calculation, the mean absolute error was ±5 units compared to a state-of-the-art CTG machine (Fig. 36b). The UCint value was smoothened using a moving average filter of width 6 to obtain the final output. We present a compact electronic sensor system to determine the MHR and UC for women in late pregnancy. Our system can give high accuracy of the maternal heartbeat from the mother's abdomen and can be further developed to capture the fetal heartbeat by enhancing the computation algorithm. The EMG module deployed with proper algorithms can also replicate the gold-standard CTG device data. Being a low-cost wearable system, compared to Doppler ultrasound and CTG, such a system can help women monitor themselves at home and reduce the trouble of visiting hospitals for a check-up. Such a system can also be deployed in primary healthcare centers in developing countries to help reduce ailments from child-birth-related complications.

# Chapter 6: Conclusion and further development

Our Maternal and Fetal heart rate monitoring device is in the present state, able to give plausibly comparable data when compared to the CTG and doppler ultrasound. With more improvisations in the further stages of the device, we will be able to match the data with higher accuracy and send it over to bluetooth. We are aiming toward doing all the processing and computing on the processor's end instead of moving to the cloud as our device has to be time-efficient in sending the data and not depend on the cloud for receiving and sending the data to and forth.

We will also employ an SD card module to our device to store the data in a stable manner and use the SPI communication protocol so that the data is not missed while sending it over through bluetooth to the phone, keeping it low energy.

Suppose this communication through SD card as well as the data accuracy is achieved over time in the subsequent versions of the prototype. In that case, we also aim to make the circuit on a flexible PCB so that it sticks to the abdomen and the patient/subject does not have to carry around the device to monitor the data.

# Publications

- A system for antenatal and intrapartum fetal monitoring with screening for fetal cardiac anomalies - Patent Filed

- S. Malkurthi, K. V. Reddy Yellakonda, A. Tiwari and A. M. Hussain, "Low-cost Color Sensor for Automating Analytical Chemistry Processes," *2021 IEEE Sensors*, 2021, pp. 1-4, doi: 10.1109/SENSORS47087.2021.9639569

- A. Navnit, D. Devendra, A. Tiwari and A. M. Hussain, "KiteCam – a novel approach to low-cost aerial surveillance," *2020 IEEE SENSORS*, 2020, pp. 1-4, doi: 10.1109/SENSORS47125.2020.9278918.

- Labour Monitoring in Pregnant Women using Electrocardiography and Electromyography Anushka Tiwari∗ , Shirley Chauhan∗ , Sailaja Bharatala† and Aftab M. Hussain∗ , Member, IEEE ∗PATRIoT Lab, International Institute of Information Technology, Hyderabad, India. – Published  IEEE APSCON 2023 - Bengaluru(Karnataka),India.